\documentclass[usenatbib,twocolumn]{mn2e}
\usepackage{amsfonts}
\usepackage{amsmath}
\usepackage{amssymb}
\usepackage{graphicx}
\usepackage{color}

\def\lsim{~\rlap{$<$}{\lower 1.0ex\hbox{$\sim$}}}
\def\bsim{~\rlap{$>$}{\lower 1.0ex\hbox{$\sim$}}}

\def\hmsun{\ {\rm M_\odot/{\it h}}}

\def\la{\langle}
\def\ra{\rangle}

\def\ln{{\rm ln}}

\def\vk{\mathrm{\bf k}}
\def\vq{\mathrm{\bf q}}

\def\vx{\mathrm{\bf x}}
\def\vy{\mathrm{\bf y}}

\def\fnl{f_\text{NL}}
\def\bng{b_\text{NG}}
\def\bpbs{b_\text{NG}^\text{\tiny pbs}}

\def\fesp{f_{\rm ESP}}

\def\bnh{\bar{n}_\text{h}}

\def\apjl{Astrophys.\ J.\ Lett.}

\def\aap{Astron.\ Astrophys.}

\def\apjs{Astrophys.\ J. Supp.}

\title[The effect of a stochastic moving barrier]
      {Scale-dependent bias from an inflationary bispectrum: the effect of a stochastic moving barrier}

\author[M. Biagetti \& V. Desjacques]
{Matteo Biagetti \& Vincent Desjacques \\  
      D\'epartement de Physique Th\'eorique and 
      Center for Astroparticle Physics (CAP), Universit\'e de Gen\`eve, \\ 
      24 quai Ernest Ansermet, CH-1211 Gen\`eve, Switzerland \\
}

\begin{document}
\pagerange{\pageref{firstpage}--\pageref{lastpage}}

\maketitle 

\label{firstpage}

\begin{abstract}

With the advent of large scale galaxy surveys, constraints on primordial non-Gaussianity 
(PNG) are expected to reach ${\cal O}(\fnl) \sim 1$. In order to fully exploit the 
potential of these future surveys, a deep theoretical understanding of the signatures 
imprinted by PNG on the large scale structure of the Universe is necessary. 
In this paper, we explore the effect of a stochastic moving barrier on the amplitude of
the non-Gaussian bias induced by local quadratic PNG. We show that, in the peak approach  
to halo clustering, the amplitude of the non-Gaussian bias will generally differ from the 
peak-background split prediction unless the barrier is flat and deterministic. 
For excursion set peaks with a square-root barrier, which reproduce reasonably well the 
linear bias $b_1$ and mass function $\bnh$ of SO haloes, the non-Gaussian bias amplitude 
is $\sim 40$\% larger than the peak-background split expectation
$\partial\ln\bnh/\partial\ln\sigma_8$ for haloes of mass $\sim 10^{13}\hmsun$ at $z=0$.
Furthermore, we argue that the effect of PNG on squeezed configurations of the halo 
bispectrum differs significantly from that predicted by standard local bias approaches. 
Our predictions can be easily confirmed, or invalidated, with N-body simulations.

\end{abstract}


\section{Introduction}
\label{sec:intro}

Upcoming large scale structure surveys will take over the hunt for primordial 
non-Gaussianity (PNG) from CMB experiments.
The recent (individual) limits on the nonlinear parameter $\fnl$ from measurements 
of galaxy clustering and the integrated Sachs-Wolfe (ISW) effect are already at 
the level of the CMB pre-Planck constraints, i.e. $\Delta\fnl\sim 80$ 
\citep{giannantonio/ross/etal:2014,ho/agarwal/etal:2013,leistedt/peiris/roth:2014}.
Forecasts for future Euclid-like galaxy surveys show that a measurement of the 
large scale galaxy power spectrum alone can constrain $\fnl\sim$ a few
\citep[e.g.][]{giannantonio/porciani/etal:2012,camera/santos/maartens:2014,
ferramacho/santos/etal:2014,deputter/dore:2014,dore/bock/etal:2014}, whereas intensity mappings of the 
21cm emission line of high-redshift galaxies could achieve $\Delta\fnl\sim 1$
\citep[e.g.][]{camera/santos/etal:2013}.

One of the most powerful large scale structure probes of PNG to date is the 
galaxy/quasar power spectrum.
In the original derivation of \cite{dalal/dore/etal:2008}, the amplitude $\bng$ 
of the scale-dependent bias induced by a primordial bispectrum of the local shape 
(i.e. $\fnl\phi^2$) was found to be proportional to the first-order bias, i.e. 
$\bng=\delta_c b_1$. 
\cite{slosar/hirata/etal:2008} used the peak-background split argument 
\citep{kaiser:1984,bardeen/bond/etal:1986} to argue 
that the amplitude of the non-Gaussian bias is proportional to the logarithmic 
derivative of the halo mass function w.r.t. $\sigma_8$ (or any proxy for the 
normalisation amplitude), i.e. $\bng=\bpbs$, where 
\begin{equation}
\label{eq:bpbs}
\bpbs\equiv\frac{\partial\ln\bnh}{\partial\ln\sigma_8} \;.
\end{equation}
\cite{scoccimarro/hui/etal:2012} generalised the peak-background split approach 
to include the non-Markovian (in the excursion set sense) and non-universality 
of the mass function. Nevertheless, they assumed that the first-crossing 
distribution is of the particular form $f(\delta_c,\sigma_0^2)$, i.e. a flat 
barrier. For $\nu_c\gg 1$, all these predictions converge to the high-peak result 
derived in \cite{matarrese/verde:2008}. Here, $\nu_c=\delta_c/\sigma_0$
is the peak significance and $\sigma_0$ is the rms variance of the mass density
field.

Both analytic models of halo collapse and numerical simulations support the fact
that, at a given halo mass $M$, the linear threshold for halo collapse is not the 
deterministic constant $\delta_c$ predicted by spherical collapse
\citep{bond/myers:1996,sheth/mo/tormen:2001,desjacques:2008a,dalal/white/etal:2008,
robertson/kravtsov/etal:2009}.
Owing to the tidal shear and other nonlinear effects, the average linear threshold 
for collapse is a monotonically increasing function of decreasing halo mass. 
Furthermore, this collapse threshold fluctuates from halo to halo because it is 
strongly sensitive to the local density and shear configuration. 
To the best of our knowledge however, the effect of a moving barrier on the amplitude 
of non-Gaussian bias has thus far been discussed only in \cite{afshordi/tolley:2008}
and \cite{adshead/baxter/etal:2012}. 
\cite{afshordi/tolley:2008} argued that the formula of \cite{dalal/dore/etal:2008} 
remains valid if one substitutes $\delta_c b_1\to \delta_{ec} b_1$, where $\delta_{ec}$ 
is the threshold for ellipsoidal collapse. 
\cite{adshead/baxter/etal:2012} investigated the effect of an ellipsoidal barrier on
the non-Gaussian bias within the path integral approach to excursion set 
\citep[see][]{maggiore/riotto:2010a}. They found that the non-Gaussian bias amplitude 
is generally different from $\delta_{ec} b_1$. However, both papers did not consider
the stochasticity of the collapse barrier.

In this paper, we will explore the effect of a realistic, stochastic moving barrier 
on the non-Gaussian bias of dark matter haloes within the peak theory framework. 
We will demonstrate that, if the (excursion set) peak theory is correct, then the 
amplitude of the non-Gaussian bias is {\it not} given by the ``peak-background split''
~\footnote{We use quotation marks here to insist on the fact that the peak-background
split really is about a change in the background density, not the normalisation
amplitude. Hence, the notation $\bpbs$ is somewhat misleading. Nevertheless, we stick
to this denomination because Eq.(\ref{eq:bpbs}) is often referred to as the 
peak-background split amplitude in the literature. We thank Ravi Sheth 
for reminding us of this important point.}
relation Eq.(\ref{eq:bpbs}). The paper is organised as follows. 
In \S\ref{sec:theory}, we explore the effect of a stochastic moving barrier on 
the non-Gaussian bias calculated in the peak approach. In \S\ref{sec:squeezed},
we discuss the implications of our findings for the squeezed limit of the galaxy
bispectrum. In \S\ref{sec:conclusion}, we conclude with a discussion about the 
validity of the peak-background split.

\section{Non-Gaussian bias with stochastic moving barrier}
\label{sec:theory}

\subsection{Stochastic barrier in Excursion set peaks}
\label{sub:esp}

\cite{sheth/mo/tormen:2001} argued that, owing to the triaxiality of the collapse, 
the linearly evolved critical density for collapse is not constant and equal to 
$\delta_c=1.68$, but rather distributed around a mean value that increases with 
decreasing halo mass. N-body simulations, which can be used to trace haloes back to
the linear density field, indeed support this prediction and indicate that the 
scatter around the mean barrier is always significant 
\citep[see e.g.][]{dalal/white/etal:2008,robertson/kravtsov/etal:2009,
ludlow/porciani:2011,ludlow/Borzyszkowski/porciani:2014}.

\begin{figure}
\center
\resizebox{0.48\textwidth}{!}{\includegraphics{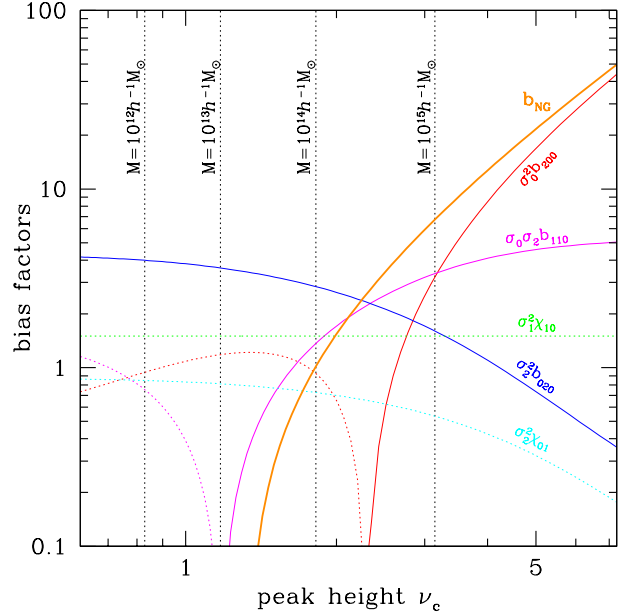}}
\caption{Dimensionless second-order bias factors in the excursion set peak approach
for the constant, deterministic barrier $B(\sigma)=\delta_c$. Dotted curves represent
negative values. $\bng$ shown as the thick solid curve is the sum of all these contributions 
(Eq.(\ref{eq:bNG})) and, according to the peak approach, is equal to the amplitude of 
the non-Gaussian bias. For clarity, we have not shown the bias factors $b_{101}$, 
$b_{011}$ and $b_{002}$ which arise from the first-crossing constraint.}
\label{fig:pbs1}
\end{figure}

Several implementations of moving and stochastic barriers exist in the literature,
ranging from direct implementations of triaxial collapse 
\citep{bond/myers:1996,sheth/mo/tormen:2001}, multidimensional excursion set approaches
\cite{sheth/tormen:2002,achitouv/rasera/etal:2013,sheth/chan/scoccimarro:2013,castorina/sheth:2013}
to the diffusive drifting barrier approach of
\cite{maggiore/riotto:2010b,corasaniti/achitouv:2011,ma/maggiore/etal:2011}. In what 
follows, we will adopt the simple prescription of
\cite{paranjape/lam/sheth:2012,paranjape/sheth/desjacques:2013}, in which $\delta_c$ 
is replaced by a generic moving barrier $B(\sigma_0)$ and the scatter is parametrised
by a random variable $\beta$. We will consider the square-root stochastic barrier 
\begin{equation}
\label{eq:barrier}
B(\sigma_0)= \delta_c + \beta \sigma_0 \;,
\end{equation}
where the stochastic variable $\beta$ closely follows a lognormal distribution with
mean $\langle \beta \rangle =0.5$ and variance Var$(\beta)=0.25$. This barrier 
furnishes a good description of the linearly extrapolated collapse threshold of SO
(Spherical Overdensity) haloes identified with a constant overdensity $\Delta_c=200$ 
relative to background \citep{robertson/kravtsov/etal:2009}. 

\cite{paranjape/sheth/desjacques:2013} interpreted this moving barrier as follows:
each halo ``sees'' a moving barrier $B=\delta_c+\beta\sigma_0$ with a value of 
$\beta$ drawn from a lognormal distribution. Here, we will adopt the interpretation
of \cite{biagetti/chan/etal:2014}, which states that each halo ``sees'' a constant 
(flat) barrier with a height that varies on an object-by-object basis. 
Consequently, the first-crossing condition does not involve any derivative of 
$B(\sigma_0)$ w.r.t. the halo mass, and we simply get
\begin{equation}
\label{eq:crossing}
B < \delta < B + \mu\, \Delta R_T \;,
\end{equation}
where $\mu=-d\delta/dR_T$ and $R_T$ is the Top-hat radius of the Lagrangian patch which 
collapses to form a halo 
\citep[see][for early implementation of the first-crossing conditions]{bond:1989,appel/jones:1990}.
Consequently, the variable 
$\mu$ will satisfy the constraint $\mu > 0$ rather than $\mu > -dB/dR_T$ as in 
\cite{paranjape/sheth/desjacques:2013}. This has a very small impact on the predicted 
halo mass function and, at the same time, simplifies the effective peak bias expansion 
since they are no correlations induced by the barrier itself (but they may be present
for actual halos).

The halo mass function predicted by the model is
\begin{align}
\label{eq:nhalo}
\frac{d \bar{n}_{\rm h}}{d\ln M} &= \frac{\bar{\rho}}{M} \nu_c \fesp(\nu_c,R_T)
\frac{d\log\nu_c}{d\log M} \\
&= -\frac{1}{3} R_T\left(\frac{\gamma_{\nu\mu}\nu_c}{\sigma_{0T}}\right) V^{-1}\fesp(\nu_c)
\nonumber \;,
\end{align}
where $\sigma_{0T}$ is the zeroth-order spectral moment of the linear density field 
smoothed with a Top-hat filter, $\gamma_{\nu\mu}$ is the cross-correlation between
the $\nu$ and $\mu$ fields and $V$ is the Lagrangian volume of a halo.
$\fesp(\nu_c)$ is the multiplicity function of the so-called excursion set peaks 
\citep{paranjape/sheth:2012,paranjape/sheth/desjacques:2013},
\begin{align}
\label{eq:fesp}
\fesp(\nu_c) &= \left(\frac{V}{V_*}\right) \frac{1}{\gamma_{\nu\mu}\nu_c}\int_0^\infty 
d\beta\, p(\beta) \\
& \quad\times \int_0^{\infty}d\mu\,\mu\int_0^{\infty} du\, f(u)\, 
\mathcal{N}(\nu_c+\beta,u,\mu) \nonumber \;.
\end{align}
Here, $V\propto R_s^3$ is the Lagrangian volume associated with the Top-Hat smoothing
 filter, $V_*  $ is the characteristic volume of peaks,  $p(\beta)$ is a log-normal 
distribution, for which we take $\langle \beta \rangle =0.5$ and Var$(\beta)=0.25$ 
as discussed above. Finally, $u$ is the peak curvature, and $f(u)$ is the peak shape 
factor of \cite{bardeen/bond/etal:1986} (BBKS). 
Clearly, the ESP mass function is not universal since $\fesp$ is a complex function
of $\nu_c$ and the spectral moments $\sigma_i$. In addition, random walks associated
with excursion set peaks are non-Markovian owing to the shape of the Top-hat and 
Gaussian filters.

\subsection{Non-Gaussian bias and peak-background split}

As shown in \cite{desjacques/gong/riotto:2013}, the non-Gaussian bias of excursion set 
peaks induced by a primordial non-Gaussianity of the form $\fnl\phi^2$, where the 
nonlinear parameter $\fnl$ is scale-independent, has an amplitude given by
\begin{align}
\label{eq:bNG}
b_{\rm NG} &= \sigma_0^2 b_{200} + 2 \sigma_1^2 b_{110}
+\sigma_2^2 b_{020}+2\sigma_1^2\chi_{10} + 2\sigma_2^2\chi_{01} \\
& \quad + \Delta_0^2 b_{002} - (\sigma_0^2)' b_{101} - (\sigma_1^2)' b_{011} 
\nonumber \;.
\end{align}
Here, a prime denote a derivative w.r.t. Top-hat radius $R_s$. $b_{ijk}$ and $\chi_{ij}$
are the ESP peak bias factors that can be derived from the ESP peak ``localised'' number
density using a peak-background split argument 
\citep[see][for details and notations]{desjacques:2013,desjacques/gong/riotto:2013}. This
is particularly interesting because the right-hand side was obtained from the ``effective''
bias expansion introduced in \cite{desjacques:2013}. In Fig.\ref{fig:pbs1}, some of the
second order bias factors together with the resulting behaviour of $\bng$ are shown for 
the constant barrier $B(\sigma_0)=\delta_c$ as a function of the peak significance $\nu_c$. 

\begin{figure}
\center
\resizebox{0.48\textwidth}{!}{\includegraphics{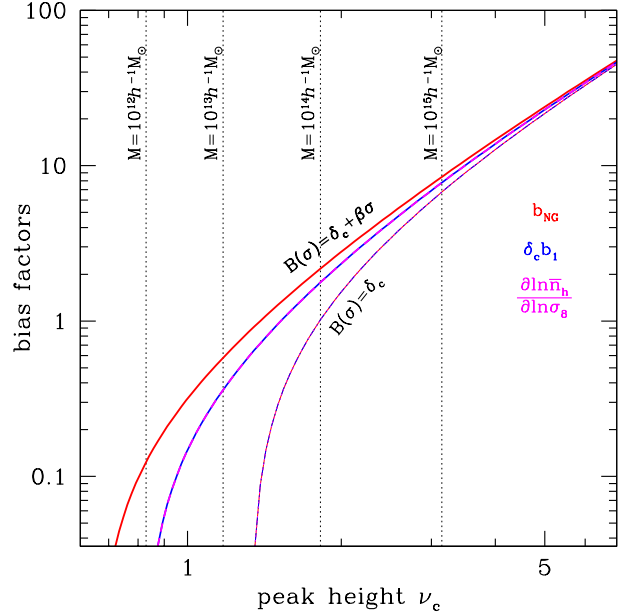}}
\caption{A comparison between the non-Gaussian bias amplitude predicted by peak theory 
($\bng$), by the peak-background split ($\partial\ln\bnh/\partial\ln\sigma_8$) and that
commonly used in forecasts ($\delta_c b_1$). All the theoretical curves were obtained 
from the excursion set peak approach assuming either a constant deterministic barrier
$B(\sigma_0)=\delta_c$ or a square-root barrier $B(\sigma_0)=\delta_c+\beta\sigma_0$,
where $\beta$ is lognormally distributed. Vertical lines indicate the corresponding 
halo masses for the fiducial $\Lambda$CDM cosmology with normalisation $\sigma_8=0.81$.}
\label{fig:pbs2}
\end{figure}

For this constant, deterministic barrier, \cite{desjacques/gong/riotto:2013} demonstrated
that the amplitude $\bng$ of the non-Gaussian bias satisfies
\begin{equation}
\label{eq:relation1}
\bng = \delta_c b_1 = \bpbs\;,
\end{equation}
where $\bnh$ is the excursion set peaks mass function Eq.(\ref{eq:nhalo}) and $b_N\equiv
b_{N00}$ is the $k$-independent piece of the $N$th-order Lagrangian, Gaussian bias (the
usual Lagrangian bias parameters in the standard local bias model). They also showed that
\begin{equation}
b_N = \frac{(-1)^N}{\bar{n}_{\rm h}}\frac{\partial^N\bar{n}_{\rm h}}{\partial\delta_c^N}\;, 
\end{equation}
in agreement with peak-background split expectations \citep{kaiser:1984}. 
Under the approximation of a constant barrier, peak theory thus predicts that the amplitude 
of the non-Gaussian bias is equally given by the sum of quadratic bias factors Eq.(\ref{eq:bNG}),
the original result $\delta_c b_1$ of \cite{dalal/dore/etal:2008} or the peak-background 
split expectation $\bpbs$ obtained by 
\cite{slosar/hirata/etal:2008}. We tested this equivalence numerically and found that it 
indeed holds. The thin, indistinguishable curves in Fig.\ref{fig:pbs2} show the various 
predictions. At this point, it is worth noticing that, although the excursion set peak 
mass function is not universal (it depends distinctly on $\delta_c$ and the spectral moments 
$\sigma_i$), the logarithmic derivative of $\bnh$ w.r.t. $\sigma_8$ is nonetheless equal 
to $\delta_c b_1$. This follows from the fact that the $\sigma_i$s conspire to appear only 
in ratios such as $\gamma_1=\sigma_1^2/(\sigma_0\sigma_2)$ or in $\nu_c=\delta_c/\sigma_0$.

Thus far however, we have followed \cite{desjacques/gong/riotto:2013} and assumed a constant 
barrier $B\equiv \delta_c$. How does the relation Eq.(\ref{eq:relation1}) change when we 
take into account the scatter and mass-dependence of the collapse barrier through the 
square-root stochastic barrier Eq.(\ref{eq:barrier}) ? To answer this question, we have 
simply computed $b_{\rm NG}$ and $b_1=b_{100}$ from the bias factors derived from the ESP 
multiplicity function Eq.(\ref{eq:fesp}). 
We have also evaluated $\bpbs$ numerically from the 
predicted halo mass function (we have again explicitly taken the numerical derivative of 
$\bar{n}_{\rm h}$ w.r.t. $\sigma_8$). The results are shown in Fig.\ref{fig:pbs2} as the 
thick solid curves. They can be summarised as follows:
\begin{equation}
\label{eq:relation2}
b_{\rm NG} \neq \delta_c b_1 = \bpbs\;.
\end{equation}
Eq.(\ref{eq:relation2}) is the main result of this paper.
While $b_{\rm NG}$ agrees with the two other quantities at the high mass end, where all 
the predictions converge towards the high-peak result of \cite{matarrese/verde:2008}, it
becomes increasingly larger as the halo mass decreases.
For the lognormal distribution of $\beta$ adopted here, deviations are quite substantial. 
Namely, for $M=10^{14}$ and $10^{13}{\it h}^{-1} M_\odot$, the predicted non-Gaussian bias
amplitude $\bng$ is $\sim 10$\% and $\sim 40$\% larger than the peak-background split 
amplitude $\bpbs$. Upon turning the scatter in $\beta$ on and off, we have found that the 
latter is driving the difference between $\bng$ and $\bpbs$ for $\nu_c\gtrsim 2$. At higher
peak heights, the discrepancy originates mainly from the fact that the barrier is not flat.

\cite{afshordi/tolley:2008} advocated the replacement $\delta_c b_1\to \delta_{ec} b_1$ to 
account for the mass-dependence of the linear collapse threshold. We have found that 
substituting $\delta_c$ either by the mean barrier $\delta_c+\sigma_0\langle\beta\rangle$ 
or by the square-root of $\langle(\delta_c+\sigma_0\beta)^2\rangle$ does improve the 
agreement with $\bng$, yet the match is far from perfect, especially around $\nu_c\sim 1$. 
Furthermore, we do not select our peaks according to their formation history. Hence, this 
has nothing to do with the assembly bias effect pointed out in \cite{slosar/hirata/etal:2008}, 
for which the extended Press-Schechter formalism of \cite{bond/cole/etal:1991} furnishes a 
good description \citep{reid/verde/etal:2010}. Finally, \cite{adshead/baxter/etal:2012}
pointed out that $\bng$ generally differs from $\delta_{ec} b_1$ (but note that they did 
not discuss the validity of $\bpbs$). However, they found a much larger effect 
than we did (see their Fig.3).
Therefore, all this strongly suggests that $\bng$ can only be written down as the sum of 
quadratic bias factors Eq.(\ref{eq:bNG}) as predicted by peak theory. 

\subsection{A closer look at the peak prediction}

For simplicity, let us momentarily ignore the variable $\mu$ as it is not essential for
understanding why a square-root stochastic barrier induces a difference between $\bng$ and
the peak-background split prediction (it is enough to retain the correlation between $\nu$
and $u$). In this case, the non-Gaussian bias amplitude takes the form
\begin{equation}
\label{eq:simplebNG}
\bng = \sigma_0^2 b_{20} + 2 \sigma_1^2 b_{11} + \sigma_2^2 b_{02} 
+2 \sigma_1^2 \chi_{10} + 2 \sigma_2^2 \chi_{01} \;.
\end{equation}
The peak bias factors $b_{ij}$ (associated with $\nu$ and $u$) and $\chi_{kl}$ (associated 
with the $\chi^2$-distributed variables) can all be computed by generalising the peak-background 
split argument to variables other than the density \citep{desjacques:2013}. In particular, 
since the peak height $\nu(\vx)$ is correlated with $u(\vx)$ (at the same position $\vx$), we 
have
\begin{equation}
\sigma_0^i\sigma_2^j b_{ij} = \frac{1}{\bar{n}_{\rm pk}}\int\!\!d^{10}\vy\,n_{\rm pk}(\vy)\,
H_{ij}(\nu,u)\,P_1(\vy)\;.
\end{equation}
Here, $n_{\rm pk}(\vy)$ is the "localised" number density of BBKS peaks (as we momentarily
ignore the first-crossing constraint), $\vy$ is a vector of 10 variables and $H_{ij}(\nu,u)$ 
are bivariate Hermite polynomials.
When stochasticity in the barrier is taken into account, $n_{\rm pk}(\vy)$ contains a 
multiplicative factor of $\delta_D\!(\nu(\vx)-\nu_c-\beta)$. 
In Eq.(\ref{eq:simplebNG}), the contribution $2\sigma_1^2\chi_{10}+2\sigma_2^2\chi_{01}$ does 
not depend on the properties of the collapse barrier because the $\chi^2$-distributed variables 
do not correlate with $\nu(\vx)$ at a given position $\vx$. Therefore, we should focus on the 
piece proportional to $b_{ij}$.

On writing the bivariate Gaussian as
\begin{equation}
\mathcal{N}(\nu,u) = 
\frac{\exp\left[-\frac{\nu^2+u^2-2\gamma_{\nu u}\nu u}{2\left(1-\gamma_{\nu u}^2\right)}\right]}
{2\pi\sqrt{1-\gamma_{\nu u}^2}}
\equiv \frac{e^{-Q(\nu,u)/2}}{2\pi\sqrt{1-\gamma_{\nu u}^2}} \;,
\end{equation}
the sum $\sigma_0^2 b_{20}+2\sigma_1^2 b_{11}+\sigma_2^2 b_{02}$ simplifies to (without writing
down the integrals over $\beta$ and $u$)
\begin{align}
\label{eq:dN8_1}
\sigma_0^2 b_{20}& + 2 \sigma_1^2 b_{11} + \sigma_2^2 b_{02} \\
& \sim \left[\frac{(\nu_c+\beta)^2+u^2-2\gamma_{\nu u}(\nu_c+\beta)u}{1-\gamma_{\nu u}^2} - 2\right]
\mathcal{N}(\nu_c+\beta,u) \nonumber \\
&\sim \Bigl(2 Q(\nu_c+\beta,u) -2 \Bigr)\mathcal{N}(\nu_c+\beta,u) \nonumber \;.
\end{align}
This should be compared to the full expression of the logarithmic derivative 
$\partial\ln\bnh/\partial\ln\sigma_8$. The latter requires evaluating derivatives of 
the multiplicity function, which is an integral of the bivariate Gaussian 
$\mathcal{N}(\nu_c+\beta,u)$ over $\beta$ and $u$ similar to Eq.(\ref{eq:fesp}). Therefore, 
the logarithmic derivative of the halo mass function w.r.t. $\sigma_8$ results in a term of the 
form
\begin{align}
\frac{\partial}{\partial\sigma_8}\mathcal{N}(\nu,u) &= 
\frac{\partial}{\partial\nu}\mathcal{N}(\nu,u)\frac{d\nu}{d\sigma_8}
+\frac{\partial}{\partial u}\mathcal{N}(\nu,u)\frac{du}{d\sigma_8} \\
&= \left[-\left(\frac{\nu-\gamma_{\nu u} u}{1-\gamma_{\nu u}^2}\right)\frac{d\nu}{d\sigma_8}
-\left(\frac{u-\gamma_{\nu u}\nu}{1-\gamma_{\nu u}^2}\right)\frac{du}{d\sigma_8}\right]
\nonumber \\
& \qquad \times \mathcal{N}(\nu,u) \nonumber \;.
\end{align}
Note that $\gamma_{\nu u}$ does not contribute since it is invariant under a (scale-independent)
rescaling of $\sigma_8$. Now, we use the fact that $\nu\equiv \nu_c+\beta$, with $\beta$
independent of $\sigma_8$ and $u\propto 1/\sigma_2$. 
Hence, $d\nu/d\sigma_8 = -\nu_c/\sigma_8$ and $du/d\sigma_8=-u/\sigma_8$. Substituting these 
derivatives in the previous expression, we arrive at
\begin{align}
\label{eq:dN8_2}
\frac{\partial}{\partial\sigma_8}\mathcal{N}(\nu_c&+\beta,u) \\
&= \frac{1}{\sigma_8}\Biggl[\frac{\left(\nu_c+\beta-\gamma_{\nu u} u\right)\nu_c 
+\bigl(u-\gamma_{\nu u}(\nu_c+\beta)\bigr)u}{1-\gamma_{\nu u}^2}\Biggr] \nonumber \\
& \qquad \times \mathcal{N}(\nu_c+\beta,u) \nonumber \;.
\end{align}
We should now compare the square brackets in Eq.(\ref{eq:dN8_1}) with that of  Eq.(\ref{eq:dN8_2}). 
We note that, in Eq.(\ref{eq:dN8_1}), there is an additional factor of $-2$ inside the brackets 
which disappears when one takes into account the first-crossing constraint. 
So, the key difference is the fact that, for $\partial\mathcal{N}/\partial\ln\sigma_8$, the square 
brackets reduce to $2 Q(\nu_c+\beta,u)$ as in Eq.(\ref{eq:dN8_2}) only if $\beta\ll \nu_c$, a 
condition which is only satisfied in the high peak limit $\nu_c \gg 1$. This is the reason why, 
in Fig.\ref{fig:pbs2}, $b_{\rm NG}$ increasingly differs from $\bpbs$ as $\nu_c$ decreases.
We also note that Eqs. (\ref{eq:dN8_1}) and (\ref{eq:dN8_2}) will differ even in the absence of 
scatter in the moving barrier (i.e. $\la\beta^2\ra=\la\beta\ra^2$). 

The peak model and peak-background split predictions will agree for a moving barrier only if 
$d\nu/d\sigma_8=-(\nu_c+\beta)/\sigma_8$ or, equivalently, if $\beta\propto \sigma_0^{-1}$. 
This implies that the deviation from $\delta_c$, $\sigma_0\beta$, does not depend on $\sigma_0$. 
However, numerical simulations \citep{sheth/tormen:2002,robertson/kravtsov/etal:2009}
clearly indicate that the scatter in the barrier increases with decreasing halo mass and is 
approximately proportional to $\sigma_0$ (hence the designation square-root barrier). Therefore, 
we shall expect $\bng\ne\bpbs$ for actual (SO) dark matter haloes if excursion set peak theory 
accurately describes their clustering properties.

\section{The squeezed limit of the galaxy bispectrum}
\label{sec:squeezed}

Retaining terms up to the fourth-point function and working within the usual local bias approximation 
$\delta_\text{h}(\vx)=b_1\delta(\vx)+(1/2)b_2\delta^2(\vx)+\dots$, the halo bispectrum with primordial
non-Gaussianity of the local type is given by 
\citep{sefusatti/komatsu:2007,sefusatti:2009,jeong/komatsu:2009}
\begin{align}
B_\text{h}(\vk_1,\vk_2,\vk_3)
&= 2 b_1^3 \biggl[\left(\fnl\frac{{\cal M}(k_3)}{{\cal M}(k_1){\cal M}(k_2)}
+F_2(\vk_1,\vk_2)\right) \nonumber \\ 
& \qquad \times P(k_1) P(k_2) \biggr] + b_2 b_1^2 P(k_1) P(k_2) \nonumber \\
& \quad + \frac{1}{2}b_1^2 b_2 \int\!\!\frac{d^3\vq}{(2\pi)^3}T(\vq,\vk_1-\vq,\vk_2,\vk_3)
\nonumber \\
& \quad + \mbox{(2 cyc.)} \;.
\end{align}
Here, ${\cal M}(k)\propto k^2$ is the transfer function between linear density and potential perturbations
and $T$ is the matter bispectrum.
We have also omitted the filtering kernels as they are not essential for the purpose of this discussion.
In the squeezed configurations, the two dominant contributions are the first and fourth term in the 
right-hand side. The first is proportional to $\fnl$ whereas the fourth contains a contribution from the 
linearly evolved primordial trispectrum proportional to $\fnl^2$, and a cross-correlation between the 
primordial bispectrum and the nonlinearly evolved density field proportional to $\fnl$.

For peaks, the analyses of \cite{desjacques:2013,desjacques/gong/riotto:2013} and the correspondence with 
the Integrated Perturbation Theory (iPT) framework \citep{matsubara:2011,matsubara:2012} indicate that the 
fourth term shall be replaced by the more general expression
\begin{equation}
\frac{1}{2} c_1(k_2)c_1(k_3) \int\!\!\frac{d^3\vq}{(2\pi)^3}c_2(\vq,\vk_1-\vq)
T(\vq,\vk_1-\vq,\vk_2,\vk_3) \;,
\end{equation}
where the linear and quadratic Lagrangian peak bias parameters $c_n(\vk_1,\dots,\vk_n)$ are given by
\begin{equation}
c_1(\vk) \equiv \left(b_{10} + b_{01} k^2\right)
\end{equation}
and
\begin{align}
c_2(\vk_1,\vk_2) &\equiv \biggl\{b_{20} + b_{11} \left(k_1^2+k_2^2\right) 
+ b_{02} k_1^2 k_2^2  \\ 
& \qquad -2 \chi_{10} \left(\vk_1\cdot\vk_2\right) +\chi_{01}
\biggl[3\left(\vk_1\cdot\vk_2\right)^2 -k_1^2 k_2^2\biggr]\biggr\} \nonumber  \;.
\end{align}
Here again, we have ignored the first-crossing constraint and omitted multiplicative factors of filtering 
kernels for sake of conciseness. Restricting ourselves to the contribution of the primordial trispectrum,
terms of the form
\begin{multline}
\fnl^2 c_1(k_2) c_1(k_3) {\cal M}(k_2){\cal M}(k_3) \Bigl[P_\phi(k_2)+P_\phi(k_3)\Bigr] \\
\times P_\phi(k_1)\int\!\!\frac{d^3\vq}{(2\pi)^3}\,{\cal M}^2(q) c_2(\vq,-\vq) P_\phi(\vq) \;.
\end{multline}
arise in the squeezed configurations $k_1\to 0$. Since ${\cal M}^2(k) P_\phi(k)\equiv P(k)$, where 
$P_\phi(k)$ is the power spectrum of the Gaussian part of the primordial curvature perturbation, the 
integral over $c_2(\vq,-\vq)$ simplifies to (after re-introducing the filtering kernels)
\begin{align}
\int\!\!\frac{d^3\vq}{(2\pi)^3}\,c_2(\vq,-\vq) P(\vq)&=
\sigma_0^2 b_{20}+2\sigma_1^2 b_{11} +\sigma_2^2 b_{02} \\ 
& \quad +2\sigma_1^2\chi_{10}+2\sigma_2^2\chi_{01} \nonumber \\
& \equiv \bng \nonumber \;.
\end{align}
Therefore, this suggests that some of the terms proportional to $\sigma_0^2 b_2$ in a calculation which 
assumes the standard local bias \citep[e.g.][]{sefusatti:2009,jeong/komatsu:2009} are, in fact, proportional 
to $\bng$. Since $\bng$ is noticeably different than $\sigma_0^2 b_2$ (see Fig.\ref{fig:pbs1}), this will of 
course have a large impact on the magnitude of the PNG signal and its dependence on halo mass. 
However, we stress again that, unless the barrier is flat and deterministic, $\bng$ cannot be replaced by the 
peak-background split expression $\bpbs$. It will also be useful to compare the predictions of the peak 
approach with e.g. the models of \cite{baldauf/seljak/senatore:2011,sefusatti/crocce/desjacques:2012}, 
which are based on the multivariate bias scheme of \cite{giannantonio/porciani:2010}. We leave all this for
future work.




\section{Conclusion}
\label{sec:conclusion}

The peak-background split has become the standard lore in analytic models of large scale structure. 
However, our findings raise concerns about its validity when it comes to the non-Gaussian bias of
actual dark matter haloes. Our analysis builds on peak theory, which furnishes a good fit to the mass
function and linear bias of SO haloes, and suggests that the peak-background split gives the wrong 
answer when the barrier is moving and stochastic. The latter is a reasonable description of the 
scatter plots $\sigma_0 - B(\sigma_0)$ constructed from numerical simulations. Based on our findings,
we predict that
\begin{itemize}
\item The non-Gaussian bias amplitude $\bng$ of SO haloes is {\it not} given by the standard 
``peak-background split'' expression, i.e.
\begin{equation}
\bng\ne\bpbs\;.
\end{equation}
The fractional departure is expected to increase with decreasing halo mass in the proportions 
shown in Fig.\ref{fig:pbs2}.
\end{itemize}
In light of the model assumptions, this inequality strictly applies to dark matter haloes closely 
related to an initial density peak, which is approximately the case for $M\gtrsim M_\star$ 
\citep{ludlow/porciani:2011}. Notwithstanding this, it will be very instructive to consider also
haloes with $M\sim M_\star$.
If the simulations turn out to support $\bng=\bpbs$ even for massive haloes, then this would
imply that either the peak approach is wrong or that moving stochastic barrier are not 
properly implemented in this framework.

We stress that our prediction is strictly valid for SO haloes only since the excursion set peak
model used in the present analysis was calibrated with SO haloes identified with a fixed 
overdensity $\Delta_c=200$ relative to the background. 
For FoF haloes for instance, the amplitude of non-Gaussian bias is suppressed relative to 
$\delta_c b_1$. We believe that this is also related to the mass-dependence and stochasticity 
of barrier. Nevertheless, we will postpone a more detailed analysis to future work since such a  
discussion is beyond the scope of this work.

Our analysis has also revealed that $\delta_c b_1=\bpbs$ even 
though the excursion set peak mass function is not universal. As we have shown, this follows 
from the fact that, in peak theory, $\bnh$ depends only on ratios of the spectral moments 
$\sigma_i$ in addition to $\nu_c=\delta_c/\sigma_0$. Note, however, that this equality may 
hold only for square-root barriers.
Furthermore, the functional dependence $\bnh(\nu_c,\gamma_{\nu\mu},\dots)$ may be very peculiar 
to the peak approach. Hence, it is unclear whether the clustering of actual dark matter haloes
satisfies $\delta_c b_1=\bpbs$. Nevertheless, it will be very instructive to also test this 
relation with N-body simulations.

Finally, \cite{desjacques/gong/riotto:2013,desjacques/crocce/etal:2010} also showed that, when 
the first-crossing condition is included, the scale-independent piece of the linear, Lagrangian 
peak bias satisfies
\begin{equation}
b_1 = -\frac{1}{\bnh}\frac{d\bnh}{d\delta_c} \;,
\end{equation}
which truly follows from a peak-background split $\delta=\delta_s+\delta_l$ \citep{kaiser:1984}.
This relation has already been tested successfully in N-body simulations (Tobias Baldauf, 
private communication). It is interesting that the peak approach predicts it from first principles 
(see \cite{schmidt/jeong/desjacques:2013} for another justification), in the sense that
$b_{10}$ was independently obtained from a calculation of the peak correlation function whereas 
the right-hand side was obtained by explicitly taking the derivative of $\bnh$ w.r.t. $\delta_c$ 
Overall, measuring separately $\bng$, $\partial\ln\bnh/\partial\ln\sigma_8$ and $\delta_c b_1$ 
will help constraining the shape of the collapse barrier.

To conclude, we note that, if $\bng>\bpbs\approx \delta_c b_1$, 
then this may at least partly explain the results of 
\cite{desjacques/seljak/iliev:2009,hamaus/seljak/desjacques:2011}, who measured a significant
increase in the amplitude of the non-Gaussian bias for SO haloes with evolved linear bias 
$b_1^\text{E}\lesssim 2$. If all this turns out to be correct, then our current forecasts for 
measurements of $\fnl$ from the non-Gaussian halo bias may be in need of revision. Ongoing work 
is aimed at testing these predictions with numerical simulations, in an attempt to (in)validate
peak theory. Finally, we also stress that these considerations apply to any tracer of the large 
scale structure whose distribution can effectively be represented by a stochastic moving barrier.

\section*{Acknowledgment}

M.B. and V.D. acknowledge support by the Swiss National Science Foundation.

\bibliographystyle{mn2e}
\bibliography{references}

\label{lastpage}

\end{document}